 \documentclass[aps, prx, reprint, superscriptaddress]{revtex4-2}
 
\usepackage{amsmath} 
\usepackage{amssymb}  

\usepackage{xcolor}         
\usepackage{latexsym}
\usepackage{amsmath}
\usepackage{amssymb}
\usepackage{times}
\usepackage{graphicx}
\usepackage{epsfig}
\usepackage{array}
\usepackage{flafter}
\usepackage{listings}
\usepackage{algorithm,algpseudocode}
\usepackage{ulem}
\usepackage{verbatim}
\usepackage{hyperref}
\usepackage{qtree}
\usepackage{mathrsfs,amsmath} 

\usepackage[a4paper, total={6in, 8.5in}]{geometry}

\usepackage{mathptmx}

\graphicspath{ {Figures/} }

\usepackage{setspace}
\usepackage{dcolumn}  
\usepackage{bm}       

\newcommand{\Ex}[1]{Example~\ref{ex:#1}}

\newtheorem{definition}{Definition}
\newtheorem{theorem}{Theorem}

\newtheorem{ExampleDef}{Example}

\newcommand{\Example}[3]{
  \begin{list}{}{
      \setlength{\leftmargin}{0em}} 
    \item                           
    \small                          
    \begin{ExampleDef} \rm          
      {\it \hspace{-1ex} #1}       
      #2                                
      \hfill {\large \boldmath $\Box$}  
      \label{ex:#3}                      
    \end{ExampleDef}
  \end{list}}

\begin{document}

\title{Computing quantum waves exactly for nonlinear potentials and curved spaces}
\author{Winfried Lohmiller}
\author{Jean-Jacques Slotine}
\affiliation{Massachusetts Institute of Technology}



\begin{abstract}

Recent work shows that the Schr\"odinger equation can be solved exactly based only on classical least action.
The computation is based on solving a Hamilton-Jacobi equation for the action, computing the classical density accordingly along all stationary action paths, and finally constructing the exact wave function based on these classical quantities alone. The method requires that the action Laplacian (or more generally the Laplacian of the propagated density) be purely time-varying along stationary action paths. In the case of arbitrary nonlinear potentials, this condition can still be verified without loss of generality by using a time rescaling.

As the complexity of the wave computation is thus shifted to that of the action and the possible time rescaling, this paper proposes a new analytical approach both to compute the action itself for a general nonlinear Hamilton-Jacobi p.d.e., and to concurrently construct a time rescaling as needed. In contrast to solving the Schr\"odinger equation directly, this computation of action and density  extends naturally to systems with nonlinear potentials or position-dependent inertia tensors. Hence, in principle it can replace the approximations of quantum perturbation theory.

For general nonlinear potentials, extending a result of Duru and Kleinert, the time-rescaling in a given metric is shown to correspond to a change of variables ${\bf q}'({\bf q})$ in the computed action, density and wave, with each eigenwave $\Psi_k$ computed using that change of variables, $\Psi_k({\bf q}) = \mathscr{H}_k({\bf q}'({\bf q}))$. Because the wave construction from classical least action paths involves only the action Laplacian, a scalar,  rather than the action Hessian, the computation of the time-rescaling is straightforward. In a $1$-dimensional system, the nonlinear oscillator transforms into a harmonic oscillator, for which an exact analytic solution is already known. For higher-dimensional systems the same approach may be used, or alternatively the $N$-dimensional Hamiltonian systems can first be decomposed into $N$ decoupled Hamiltonian dynamics. While for purely time-dependent action Laplacians time-rescaling is never needed, it may facilitate the computation of the classical action itself.

We show that the approach makes it straightforward to construct the quantum wave for basic cases where no exact solution has been yet derived, with the time rescaling unifying the computation. 
The approach, first illustrated  for the known three dimensional hyperbolic potential waves of the hydrogen atom, is then used to compute the quantum  waves for 
a quartic oscillator, for which so far no direct exact solutions are known.



\end{abstract}

\maketitle

\section{Introduction}

Richard Feynman's path-integral formulation of quantum mechanics~\cite{Feynman48} is a central tool among results aimed 
at bridging classical and quantum physics. Recent work \cite{LohmillerSlotine2026QuantumWaves} shows that the Schr\"odinger equation can be solved exactly based only on classical least action, in essence replacing all the zig-zag paths in Feynman's construction by a much sparser set of least action paths, appropriately weighted according to the classical density along each path.  The question of efficient wave computation is thus shifted to that of efficient action computation, which is the subject of this paper. In turn, this opens the possibility of exact wave solutions for quantum systems with nonlinear potentials or position-dependent inertia tensors. For general nonlinear potentials, the construction in \cite{LohmillerSlotine2026QuantumWaves} may require a time-rescaling similar to \cite{Duru}, as it assumes that along stationary action paths the density is only a function of the rescaled time. Along with the action computation, this paper details the derivation of this time-rescaling when needed, concurrently with the implied change of position variables ${\bf q}'({\bf q})$ in the corresponding eigenwaves.

Recall that the classical motion of a physical system corresponds to a local extremum over variational paths ${\bf q} = (q^1, ..., q^N) \in \mathbb{R}^N$ of the system's action,  
\begin{eqnarray}
    \phi({\bf q}, t, {\bf q}_o \veebar {\bf p}_o)  &=& \underset{{\bf q}(t)}{\rm{extremum}} 
   \int L(\dot{\bf q}(t), {\bf q}(t),t) \ d t \label{eq:Lagrangian} \\
   L &=& \frac{1}{2} \ \dot{\bf q}^T {\bf M} \dot{\bf q} +  {\bf A}^T {\bf Q} \ \dot{\bf q}  - V 
   \label{eq:action}  \nonumber
\end{eqnarray}
with final position ${\bf q}$ at time $t$, initial position ${\bf q}_o$ or (exclusive $\veebar$) initial momentum ${\bf p}_o \ $,  Lagrangian $L$, inertia tensor or metric ${\bf M}({\bf q})$, potential energy $V({\bf q})$ (which has an ambiguity w.r.t. a constant energy offset $E$), vector potential ${\bf A}({\bf q})$, diagonal charge ${\bf Q}$ matrix with the constant charge $Q$ of each particle on the diagonal, see e.g. \cite{Feynman, Lagrange, Lovelock}. In this paper, unless otherwise specified, we will simply use the term action to refer to local stationary action. 

The action $\phi$ can be computed from $\frac{d \phi}{dt } =  L$ along a  path
with $ H \ $  piecewise  ${\bf C}^2({\bf q},t, \nabla \phi)$,
\begin{eqnarray}
 - {\frac {\partial \phi}{\partial t}}  &=&  H   \label{eq:HJ}  \\
&=&  {\frac {1}{2}} \left( \frac{\partial \phi}{\partial {\bf q}} - {\bf Q}  {\bf A} \right)^T  {\bf M}^{-1}  \left( \frac{\partial \phi}{\partial {\bf q}} - {\bf Q}  {\bf A} \right) + V  \nonumber 
\end{eqnarray}
The related multi-value action computation is summarized in section \ref{col} ~\cite{LohmillerSlotine2026QuantumWaves}.

The symmetric metric ${\bf M}({\bf q})$ is required to be uniformly invertible, but is not necessarily positive 
definite. We use the standard Laplace-Beltrami tensor operator ~\cite{beltrami1902ricerche, laplace1799mecanique, Lovelock}
\begin{eqnarray}
\nabla_{\bf M} \cdot {\bf f} &=& {\frac {1}{\sqrt {\det {\bf M} }}} \sum_{n=1}^N \frac{\partial }{\partial q^n} \left({\sqrt {\det {\bf M} }} \  f^n  \right)  \nonumber \\
 \Delta_{\bf M}  \ f &=& \nabla_{\bf M} \cdot \left(  {\bf M}^{-1} \frac{\partial f}{\partial {\bf q}}  \right) \label{eq:Delta} 
\end{eqnarray}
for a given metric ${\bf M}({\bf q})$, vector ${\bf f}({\bf q}, t) = (f^1, ..., f^N) \in \mathbb{R}^N$ and scalar $f({\bf q}, t) \in \mathbb{R} $. No subscript is used for ${\bf M}({\bf q}) = {\bf I}$. The superscript $\nabla_{\bf M}'$ is used if the coordinates are ${\bf q}'$ rather then ${\bf q}$. The vector potential ${\bf A}({\bf q},t)$ is assumed to follow the Coulomb or Lorenz gauge $\nabla_{\bf M} \cdot {\bf A} = 0 \ $ \cite{feynmanquantum1998}.

A key point is that the classical density $ \rho$  can be computed analytically along each action or momentum branch simply by integrating the classical Euler continuity o.d.e. \cite{Euler} 
\begin{eqnarray}
    0 &=& \frac{d \rho_j }{dt}  + \rho_j \  \Delta_{{\bf M}}  \phi_j  \label{eq:continuity} 
\end{eqnarray} 
A spatially independent propagated density $\rho(t)$ can be assured by a spatially independent $\Delta_{{\bf M}} \phi_j(t)$. For a spatially dependent $\Delta_{{\bf M}} \phi_j({\bf q}, t)$, making a time transformation $t_j'(t, {\bf q}) =\int_o^t T ({\bf q}(\theta), \theta) d \theta $ 
as in \cite{Duru} replaces the continuity o.d.e. (\ref{eq:continuity})  above by 
$$0 = \frac{d \rho_j }{dt'}  + \rho_j \  T \Delta_{{\bf M}}  \phi_j$$ 
A  spatially independent propagated density $\rho(t')$ can then be assured by making $T \Delta_{{\bf M}} \phi_j(t')$ spatially independent. Since  $T \Delta_{{\bf M}} \phi_j(t')$ is scalar, the existence of a suitable $T({\bf q}, t)$ is always guaranteed. This spatial independence in turn implies the absence of Bohm-like quantum potentials \cite{Bohm1952I, Bohm1952II}. Thus, the action does not have to be quadratic for the result to be exact, in contrast to the semi-classical derivations in \cite{VanVleck1928} or \cite{Feynman}.


As summarized in Section~\ref{quan} and detailed in \cite{LohmillerSlotine2026QuantumWaves, LohmillerSlotine2026BohmPotential}, Feynman's infinity of non-classical zig-zag paths can now be reduced to the subset of all extremal classical action paths, each weighted with the classical  propagated {\it spatial independent} density computed along each path. This yields an exact construction of the kernel $K'(t',{\bf q}) $ of the time-rescaled Schr\"odinger equation
{\scriptsize
\begin{equation}
\left[ \frac{\hbar}{i} \frac{\partial}{\partial t'} + \frac{1}{2}\left( \frac{\hbar}{i} \nabla_{{\bf M}T} - {\bf Q} {\bf A} \right) \cdot ({\bf M}T)^{-1} \left( \frac{\hbar}{i} \nabla -  {\bf Q}  {\bf A} \right) + \frac{V}{T} \right]  K' = 0 \label{eq:Schrodtime} 
\end{equation}}
It was shown in \cite{Duru} (equation (16)) that $K'(t',{\bf q})$ has the same eigenwaves as $\psi = \sum_k \Psi_k ({\bf q}) e^{\frac{i}{\hbar} E_kt}$, which also solves the original Schr\"odinger equation
{\small
\begin{equation}
\left[ \frac{\hbar}{i} \frac{\partial}{\partial t} + \frac{1}{2}\left( \frac{\hbar}{i} \nabla_{\bf M} - {\bf Q} {\bf A} \right) \cdot {\bf M}^{-1} \left( \frac{\hbar}{i} \nabla -  {\bf Q}  {\bf A} \right) + V  \right] \psi = 0 \label{eq:Schroed} 
\end{equation}}
This fact was exploited in \cite{LohmillerSlotine2026QuantumWaves, LohmillerSlotine2026BohmPotential} and will be detailed in Section~\ref{quan}.  

The goal of section \ref{nonlinearaction} of this paper is to systematically address the computation of the propagated action, propagated density for non-linear $V ({\bf q})$ where $\Delta_{\bf M} \phi ({\bf q},t)$ {\it is indeed space dependent}. This action computation is again made with the proposed time-rescaling since it simplifies the related computations. 
For one-dimensional systems, the proposed multi-valued action approach with time-rescaling leads to a quadratic action branch in scaled time, which can be solved exactly. This is a key improvement over approximate perturbation theory \cite{Heisenberg1925, Schrodinger1926, Wilson1929}, using the WKB approximation \cite{Wentzel1926, Kramers1926,Brillouin1926} of partial differential equations, or semi-classical approximations \cite{landauquantum, Maslov1972}. 
In general, $N>1$ dimensional systems first have to be decomposed  into $N$ separate dynamics using separation of variables or more generally the modular decomposition of nonlinear Hamiltonians recently derived in \cite{lohmiller2025contraction20naturalmetrics}. 

The approach is illustrated with several nonlinear examples in Section \ref{examples}. Concluding remarks and perspectives are offered in Section~\ref{concluding}.

\section{Computing waves exactly from action} \label{secton2}

Rather than being based on an approximation or a mathematical expansion, our mapping in~\cite{LohmillerSlotine2026QuantumWaves} from classical to quantum mechanics relies on a different and exact mechanism. This section summarizes this construction, with the reader referred to \cite{LohmillerSlotine2026QuantumWaves}  for details.

\subsection{\label{col} Multi-valued local least action and multipaths} 


Let us first introduce spatial inequality constraints on a multiply connected manifold, and then define action branches on this manifold.
\begin{definition} 
{\it The constrained multiply connected manifold $ \ \mathbb{G}^N({\bf q} \ \veebar \ {\bf p}, t) \subseteq \mathbb{R}^N$ is defined by the $g=1, ..., G$ inequality constraints $ \ f_g({\bf q}, t)\le 0\ \  \veebar \ \ f_g({\bf p}, t)  \le  0 \ $.
The set of active constraints $\mathbb{G}  \subseteq  \{1, ..., G \}$  is the set of indices $g$  on the boundary $\partial \mathbb{G}^N$ of $\mathbb{G}^N$ such that $ \ f_g({\bf q}, t) =0 \ \ \veebar \ \ f_g({\bf p}, t) = 0 \ $.}
 \label{def:equalconstraint}
\end{definition}

A local least action solution satisfies
\begin{equation}
   \frac{d}{dt} \frac{\partial L}{\partial \dot{\bf q}} - \frac{\partial L}{\partial {\bf q}} \ = \ \frac{d \nabla \phi}{dt} + \frac{\partial H}  {\partial {\bf q}} \ = \ \sum_{g \in \mathbb{G}} \lambda_g \frac{\partial f_g}{\partial {\bf q}} \label{eq:EulerLangrange}
\end{equation}
where the Lagrange parameter $\lambda_g$ defines the magnitude of the cost gradient at the active constraint.

\begin{definition}
{\it The action branch set is the set of local extremal action fields $ \phi_j({\bf q}, t), 
\ {\bf q} \in \mathbb{G}^N, t \ge 0$ which are different  at least at one $({\bf q}, t)  \ $, except for the integration constant for each given initial condition in (\ref{eq:action}).}
\label{def:branchset}
\end{definition}

Multi-valued least action branches naturally arise from multiple initial conditions in ${\bf q}_o$ or ${\bf p}_o \ $ in (\ref{eq:action}).  After the initialization, they can also arise at branch points. 
\begin{definition}
{\it The set of branch points $\mathbb{B}^N\subseteq  \mathbb{G}^N$ consists of all points where the set of distinct  $ \ \phi_j({\bf q}, t)$ changes, for $t \ge 0 \ $. Each branch point ${\bf q}$ occurs $(b_{\bf q}-1)$ times in $\mathbb{B}^N$, where $ \ b_{\bf q} \in \mathbb{N} \ $ is the number of distinct action solutions in the local neighborhood of ${\bf q}$.}
\label{def:branch}
\end{definition}

Introducing for generality an ensemble $\mathbb{E}$ of possible initial density conditions to describe different initial conditions occurring with given probabilities, \cite{LohmillerSlotine2026QuantumWaves} shows the following result. 
\begin{theorem}
The  multi-valued least action field $ \ \phi_j({\bf q}, t, {\bf q}_o \veebar {\bf p}_o)$  of the Hamilton-Jacobi p.d.e (\ref{eq:HJ}) with initial initial $\phi_{o j}({\bf q}, 0, {\bf q}_o \veebar {\bf p}_o)$ locally extremizes (\ref{eq:Lagrangian}) on the constrained multiply connected manifold ${\bf q}, {\bf p} \in \mathbb{G}^N$ of Definition \ref{def:equalconstraint}, yielding the multipaths
\begin{eqnarray}
   \frac{d \nabla \phi_j}{dt}  +  \frac{\partial H}  {\partial {\bf q}}\  &=&  \sum_{g \in \mathbb{G}} \frac{\partial f_g}{\partial {\bf q}} \lambda_g \label{eq:dotp} \\
   {\bf M}({\bf q})\ \frac{d{\bf q}}{dt}  &=& \nabla \phi_j- {\bf Q} \ {\bf A} \label{eq:dotq} 
\end{eqnarray}
where the reflection force $\lambda_g \ $ fulfills Definition \ref{def:equalconstraint}. 

The action is said to be $\mathbb{B}$-valued, where $\mathbb{B}$ indexes the set $\mathbb{B}^N$ which accounts for all branch points $\mathbb{B}^N$ of Definition \ref{def:branch}. Branch points exist only at unbounded  $\Delta_{\bf M} \phi_j({\bf q} \in \mathbb{G}^N, t)$ or unbounded $\nabla \phi_j({\bf q} \in \partial \mathbb{G}^N, t)$.

Combining (\ref{eq:continuity}), (\ref{eq:dotq}) the propagated classical density can be computed along each extremal path ${\bf q}_j(t)$, yielding the path integral 
\begin{eqnarray}
      \rho_j^{\epsilon} (t', {\bf q}_o \veebar {\bf p}_o) ) &=& p^{\epsilon}  e^{- \int_o^{t'} (T \Delta_{{\bf M}} \phi_j) (\theta')   d \theta'}   \label{eq:densitydash}
\end{eqnarray}
where the gauge $ \ \nabla_{\bf M} \cdot {\bf A} = 0 $ is used for an ensemble $\mathbb{E} \ $,  with $\sum_{\epsilon \in \mathbb{E}} p^{\epsilon}  = 1$. 

A time scaling is introduced to assure that  $(T \Delta_{{\bf M}} \phi_j)(t')$ is position independent,
\begin{equation}
    t_j'(t, {\bf q}) =\int_o^t T(\theta, {\bf q}(\theta)) d \theta \iff t(t_j', {\bf q}) = \int_o^{t'} 1 / T d \theta \label{eq:TimeScaling}
\end{equation}
Note $T=1$ if $\Delta_{{\bf M}} \phi_j$ is already position independent. 
\label{th:Hamilton}
\end{theorem}
Note that branch points in Theorem \ref{th:Hamilton} may occur at the initial condition, at singularities, or at constraints. At a branch point the momentum is undefined, which implies a stochastic event at that point. Classical physics is deterministic except at branch points, where it is stochastic, a result which allows the bridge to quantum physics in the next section.


\subsection{Exact wave computation from classical multi-valued action and density} \label{quan}

The time-rescaled Schr\"odinger equation (\ref{eq:Schrodtime}) 
can be solved on each stationary branch $j \in \mathbb{B}$ of Definition (\ref{def:branchset}) by 
the propagator or the Feynman kernel \cite{Feynman} 
\begin{eqnarray}
    K'({\bf q}, t', {\bf q}_o \veebar {\bf p}_o) &=& \sum_{j \in \mathbb{B}} \sqrt{\rho_j} \ e^{\frac{i }{\hbar} \phi_j} \label{eq:propagator} \\
    &=&
     \sum_{j \in \mathbb{B}} e^{- \frac{1}{2} \int_{0}^{t'_j} (T \Delta_{{\bf M}} \phi_j) (\theta') \ d \theta' }  e^{\frac{i }{\hbar} \phi_j} \nonumber 
\end{eqnarray}
using the action field $\phi_j({\bf q}, t(t', {\bf q}), {\bf q}_o \veebar {\bf p}_o)$ and the classical space independent density path integral (\ref{eq:densitydash}) of Theorem \ref{th:Hamilton} using the proofs of \cite{LohmillerSlotine2026QuantumWaves} or \cite{LohmillerSlotine2026BohmPotential}. No Bohm quantum potential is needed due to the spatial independence of the density (\ref{eq:densitydash}). We now have to show how the eigenwaves of the kernel (\ref{eq:propagator}) in time $t'$ correspond to the eigenwaves of the original Schr\"odinger equation (\ref{eq:Schroed}) using the main logic of \cite{Duru}: 

Weighting and integrating the propagator (\ref{eq:propagator}) with the initial orthonormal wave $\mathscr{H}_{ko}({\bf q}) =   \sqrt{\rho_{ok}}({\bf q}) \ e^{\frac{i }{\hbar} \phi_j(0, {\bf q}) }$ leads to the general eigenwave \cite{Feynman}  
\begin{equation}
\mathscr{H}_k({\bf q}) = \int_{{\bf q}'_o \ \veebar \ {\bf p}_o \in \mathbb{G}^N} \mathscr{H}_{ko} \ K' \ d{\bf q} \ \veebar \ d{\bf p} \nonumber
\end{equation}
of the time-rescaled Schr\"odinger equation (\ref{eq:Schrodtime}). In order to remove the dependence of the Hamilton-Jacobi (\ref{eq:HJ}) p.d.e. on $t$ or $t'$, we include a constant energy offset $E_k$ or $E_k'$ in the potential energy $V$ or time-scaled potential energy $V'=V / T$ of the original and time-scaled Hamiltonians,
{\scriptsize
\begin{eqnarray}
     0  &=&  H' = \frac{H}{T} = {\frac {1}{2}} \left( \frac{\partial \phi}{\partial {\bf q}} - {\bf Q}  {\bf A} \right)^T ({\bf M}T)^{-1}  \left( \frac{\partial \phi}{\partial {\bf q}} - {\bf Q}  {\bf A} \right) + V' \label{eq:ScaledTimeIndepenent} \\
    0  &=&  H =  {\frac {1}{2}} \left( \frac{\partial \phi}{\partial {\bf q}} - {\bf Q}  {\bf A} \right)^T  {\bf M}^{-1}  \left( \frac{\partial \phi}{\partial {\bf q}} - {\bf Q}  {\bf A} \right) + V  \label{eq:Indepenent} 
\end{eqnarray}}
In turn, this removes the $t'$-dependent pre-factor of the eigenwave wave $\mathscr{H}_k({\bf q}')$. 

The Laplace-Beltrami operator of the Schr\"odinger equations (\ref{eq:Schrodtime}) and  (\ref{eq:Schroed}) with the Hamiltonians (\ref{eq:ScaledTimeIndepenent}) and (\ref{eq:Indepenent}) is constrained
either to a manifold of constant $t$ or to a manifold of constant $t'$, which explains the different metrics ${\bf M}$ and ${\bf M} T$ \cite{Lovelock}. 
Equation (16) in \cite{Duru}  and \cite{ HenneauxTeitelboim1992} show that both time independent eigenwaves are the same which was exploited in~\cite{LohmillerSlotine2026QuantumWaves}.

Let us summarize the above~\cite{Duru,LohmillerSlotine2026QuantumWaves,LohmillerSlotine2026BohmPotential}.
\begin{theorem} 
The  Lagrangian kernel of the time-independent Schr\"odinger equation  (\ref{eq:Schroed}) 
{\small 
\begin{equation}
\left[ \frac{\hbar}{i}  \frac{\partial }{\partial t} + \frac{1}{2}\left( \frac{\hbar}{i} \nabla_{\bf M} - {\bf Q} {\bf A} \right) \cdot {\bf M}^{-1} \left( \frac{\hbar}{i} \nabla -  {\bf Q}  {\bf A} \right) + V  \right]  \Psi_k = 0 \label{eq:Schroedth}
\end{equation}}
can be computed from the multi-valued least action field $\phi_j({\bf q}, t(t', {\bf q}), {\bf q}_o \veebar {\bf p}_o)$ and the classical density (\ref{eq:densitydash}) of the ensemble $\mathbb{E} \ $ of Theorem \ref{th:Hamilton} with 
\begin{eqnarray}
  K'({\bf q}, t', {\bf q}_o \veebar {\bf p}_o) =
     \sum_{j \in \mathbb{B}} e^{- \frac{1}{2} \int_{0}^{t'_j} (T \Delta_{{\bf M}} \phi_j) (\theta') \ d \theta' }  e^{\frac{i }{\hbar} \phi_j} \ \ \ \ \ \ \ \ \label{eq:Wave}
\end{eqnarray}
A time scaling (\ref{eq:TimeScaling}) is introduced if the Laplacian of the propagated density $\Delta_{{\bf M}} \sqrt{\rho_j}$ is position dependent. This time scaling is not needed if $\Delta_{{\bf M}} \sqrt{\rho_j}$ is position independent, which is a weaker condition than requiring the Laplacian $\Delta_{{\bf M}} \phi_j$ to be position independent, which is itself weaker than requiring a quadratic action.

Weighting and integrating the propagator (\ref{eq:Wave}) with the initial orthonormal wave $\mathscr{H}_{ko}({\bf q}) =   \sqrt{\rho_{ok}}({\bf q}) \ e^{\frac{i }{\hbar} \phi_j(0, {\bf q}) }$ yields the general eigenwave
\begin{equation}
\mathscr{H}_k({\bf q}) = \int_{{\bf q}'_o \ \veebar \ {\bf p}_o \in \mathbb{G}^N} \mathscr{H}_{ko} \ K' \ d{\bf q} \ \veebar \ d{\bf p} 
\end{equation}
of the original Schr\"odinger equation (\ref{eq:Schroedth}). We select the arbitrary constant energy offset $E_k'$ in the definition of the time-scaled potential energy $V / T$ to remove the time-dependent pre-factor of the eigenwave wave $\mathscr{H}_k({\bf q})$.

The associated  quantum density matrix at time $t$
\begin{equation}
\varrho ({\bf q}, t) \ =  \ \sum_{\epsilon \in \mathbb{E}} p^{\epsilon} \ \psi^{\epsilon} \psi^{\epsilon \dagger} \ \ \  \label{eq:Prob}
\end{equation}
is the determined forward mapping along all classical paths of Theorem \ref{th:Hamilton} from the initial quantum density distribution at $t=0$. 
\label{th:quantum}
\end{theorem}

Note that Theorems \ref{th:Hamilton} and \ref{th:quantum} immediately extend 
to complex or quaternion actions, as long as a real classical path can be constructed, for instance using superposition. Equation (\ref{eq:Wave}) of Theorem \ref{th:quantum}
\begin{itemize}
    \item uses only the $\mathbb{B}$-valued classical multipaths or actions from Theorem \ref{th:Hamilton}, which are a subset of all zig-zag paths in Feynman's path integral 
    \begin{equation}
        K({\bf q}_o, {\bf q}, t)= \frac{1}{Z} \int_{{\bf q}_o}^{\bf q} e^{\frac{i  }{\hbar} \int_o^t L d \theta} \ \mathcal{D} {\bf q} \label{eq:Feynman}
    \end{equation}
    where $\mathcal{D} {\bf q}$ denotes the integration over $\infty^{\infty}$ stochastically time-sliced zig-zag paths and $Z$ is the normalization factor~\cite{Feynman}.  
    \item uses the same time-rescaling (15) and (16) \cite{Duru,Kleinert2009s} which was needed to apply (\ref{eq:Feynman}) to more complicated Hamiltonians as the hydrogen atom. 
    \item is very different from the Madelung density and action \cite{Madelung1927} 
\begin{equation}
    \rho = |\psi|^2 \ \ \ \ \ \ \ \ \ \ \ \ \phi = \hbar\,\arg(\psi) \nonumber
\end{equation}
which, in contrast to Theorem \ref{th:quantum}, generates for most waves a Bohm quantum potential \cite{Bohm1952I, Bohm1952II} since the Madelung density is tyically position-dependent.

\end{itemize}

The propagated density $\rho_j $ is spatially independent if $\Delta_{{\bf M}} \phi_j$ is spatially independent. This condition is immediately verified in the following cases, with the first case (only) corresponding to a well-know result of Feynman~\cite{Feynman} and Van Vleck~\cite{VanVleck1928}.

\begin{enumerate}
    \item If the action is single-valued and quadratic, as for a free particle or harmonic oscillator, then the propagated density is spatially independent, and the solution in~\cite{LohmillerSlotine2026QuantumWaves} is the same as that derived exactly along classical paths in \cite{Feynman} and \cite{VanVleck1928}.
    
    \item If the action is multi-valued and each action branch is quadratic or linear \cite{LohmillerSlotine2026QuantumWaves}, as e.g. for a potential well, tunneling, Dirac, Pauli or Maxwell spin, then the propagated density on each branch is also spatially independent. This was first introduced in \cite{LohmillerSlotine2026QuantumWaves} thanks to the exact classical branch point definition from unbounded $\Delta_{{\bf M}} \phi_j$, the classical quantization lemma 3.4, and the use of complex or quaternion actions. In tunneling for instance, the relevant action cannot be real as there is no classical correspondent, but using a complex action solves the problem easily~\cite{LohmillerSlotine2026QuantumWaves}. Similarly, spin can be described using quaternion-based action~\cite{LohmillerSlotine2026QuantumWaves}, without requiring extensions of Feynman's zig-zag paths~\cite{Schulman1968PathIntegralSpin,Nielsen1988PathIntegralQuantizeSpin}.
    \item Non-quadratic action branches as for the conic action of the two-slit experiment and the nonlinear action of the Aharonov-Bohm experiment in \cite{LohmillerSlotine2026QuantumWaves}. These cannot be assessed exactly with the semi-classical results of \cite{VanVleck1928} or \cite{Feynman}. For the hydrogen atom in~\cite{LohmillerSlotine2026QuantumWaves}, the time scaling (\ref{eq:TimeScaling}) above yields a spatially independent density
    from the nonlinear Kepler action.
    
\end{enumerate}
The goal of the next section is to systematically compute the corresponding action, density and wave for the last case, i.e. for general nonlinear actions.



\section{Nonlinear actions} \label{nonlinearaction}



Similarly to Section \ref{quan} we first transform without loss of generality the Hamilton-Jacobi equation (\ref{eq:HJ}) in a higher embedding space where ${\bf M}$ is mapped to an identity matrix times $T({\bf q})$ 
with the eikonal equation 
\begin{equation}
\frac{\partial {\bf q}'_k}{\partial {\bf q}}^T \frac{\partial {\bf q}'_k}{\partial {\bf q}} = \frac{{\bf M} }{m} T_k \ \ \ \ \ \ {\rm where} \ \ \ T_K({\bf q}) = \frac{V({\bf q})}{V'({\bf q})}
\nonumber
\end{equation}
of the Nash-Kuiper theorem \cite{Nash} and where the mass $m$ turns the inertia tensor ${\bf M}$ in a metric tensor.
\begin{itemize}
    \item to compute an analytic action solution
    \item to assure that $\Delta_{\bf M} \phi T_k$ and hence the propagated density $\rho_j(t')$ (\ref{eq:densitydash}) is space-independent
\end{itemize}
Note that the existence of the embedding coordinates $\bar{\bf q}_k({\bf q})$ is given by the Nash-Kuiper theorem \cite{Nash}, where the dimension of ${\bf q}'$ is {at most $\ M  \ge N+1\ $.

While solving the above for $N>1$ can imply numerics and singularities, Theorem \ref{th:quantum} does not necessarily mandate a quadratic action for $N>1$. Alternatively, one could first classically decouple the Hamilton-Jacobi dynamics into $N$ decoupled dynamics using separation of variables, or use the recently developed modular decomposition of nonlinear Hamiltonians derived in \cite{lohmiller2025contraction20naturalmetrics}.

A possible choice of a time-rescaled Hamiltonian is the harmonic oscillator in ${\bf q}'_k$ coordinates with a scaled potential energy $V' = \frac{\omega^2}{2m}  {\bf q}_k' {\bf q}^{'T}_k - E_k'$ and the eigenvalues $ \ E_k' = \hbar \omega (k+\frac{1}{2}),\ k \in \mathbb{N} \ $. The propagated action and density is then given by \cite{Feynman}
{\small \begin{eqnarray}
     \phi &=& \int_{z_o}^z  Q A'_n + \frac{m \omega}{2} \left( \cot{\omega t'} \ (q_n^{'2} + q_{no}^{'2} )  - \frac{2}{\sin{\omega t'}}  \ q_n' q_{ko}'   \right) \nonumber \\
 \sqrt{\rho'} &=& \sqrt{\frac{ m \omega}{2 \pi i \hbar \sin{\omega t'} }} \label{eq:AnalyticAction}
\end{eqnarray}}
where $q_n'$ is an element of the $n=1, ...N$-dimensional vector ${\bf q}'_k$. Theorem \ref{th:quantum} implies the Lagrangian kernel
\begin{eqnarray}
&& K'(q_n, t', q_{n o} )  =  \sqrt{\rho }   e^{\frac{i }{\hbar} \phi  } = \label{eq:kernel} \\
 &&  \sum_{k \in \mathbb{N}} e^{-\frac{i}{\hbar} E_k t'} \  \mathscr{H}_k(q_n') \mathscr{H}_k(q_{n o}')   \nonumber
\end{eqnarray}
with the orthonormal eigenwaves 
\begin{equation}
\mathscr{H}_k(q_n') = \sqrt[4]{\frac{M  \omega}{\pi \hbar}} \frac{1}{\sqrt{2^{k} k!}} \ H_{k} (q_n') e^{-\frac{1}{2} q_n^{'2}} \nonumber
\end{equation}
of the physicist Hermite polynomials $H_k(q') = \left( 2 q' - \frac{d}{d q'} \right)^k \cdot 1 \ $ with $k \in \mathbb{N}\ $ \cite{Feynman}.
Feynman showed this equivalence of the harmonic oscillator action to the harmonic oscillator eigenwaves in section 8.1 of \cite{Feynman}, first with a Taylor series expansion, and then alternatively with the definition of the generating function of the Hermite polynomials. In this proof, he was using neither existing quantum results, nor zig-zag paths.


Let us now summarize the above with Theorem \ref{th:Hamilton} and Theorem \ref{th:quantum}:
\begin{theorem} 
Consider the multi-valued least action field $\phi_j, j \in \mathbb{B}$ of the piecewise ${C}^2(q,t, \nabla \phi)$ Hamilton-Jacobi p.d.e. (\ref{eq:HJ}), which fulfills without loss of generality the Hamilton-Jacobi p.d.e
\begin{eqnarray}
 - {\frac {\partial \phi}{\partial t}}  &=&  H   \label{eq:HJscaled}  \\
&=&  {\frac {1}{2}} \left( \frac{\partial \phi}{\partial {\bf q}} - {\bf Q}  {\bf A} \right)^T {\bf M}^{-1} \left( \frac{\partial \phi}{\partial {\bf q}} - {\bf Q}  {\bf A} \right) + V  \nonumber 
\end{eqnarray}
of Theorem \ref{th:Hamilton}. 
The change of variables 
\begin{equation}
\frac{\partial {\bf q}'_k}{\partial {\bf q}}^T \frac{\partial {\bf q}'_k}{\partial {\bf q}} = \frac{{\bf M} }{m} T_k  \ \ \ \ \ \ {\rm where} \ \ \ T_K({\bf q}) = \frac{V({\bf q})}{V'({\bf q})}
\label{eq:ChangeOfVariables}
\end{equation}
with $V' = \frac{m}{2} \omega^2 {\bf q}_k^T {\bf q}^{'T}_k - E_k'$ and eigenvalues $E_k' = \hbar \omega (k+\frac{1}{2}), k \in \mathbb{N}$ leads to a quadratic action.

The classical propagated action and density from ${\bf q}_o'({\bf q}_o)$ is given by (\ref{eq:AnalyticAction}). This implies according to Theorem \ref{th:quantum} the Lagrangian kernel (\ref{eq:kernel}).

Finally the eigenwave 
\begin{eqnarray}
    && \Psi_k(q_n) = \mathscr{H}_k(q_n'(q_n)) = \nonumber \\ 
    &&  \sqrt[4]{\frac{M  \omega}{\pi \hbar}} \frac{1}{\sqrt{2^{k} k!}} \ H_{k} (q_n') e^{-\frac{1}{2} q_n^{'2}} \label{eq:EigenWave}
\end{eqnarray}
where $q_n'$ is an element of the $n=1, ...N$-dimensional vector ${\bf q}'_k$,
solves the original time-independent Schr\"odinger equation (\ref{eq:Schroedth}) 
{\footnotesize 
\begin{eqnarray}
   && \left[  \frac{1}{2}\left( \frac{\hbar}{i} \nabla_{\bf M} - {\bf Q} {\bf A} \right) \cdot {\bf M}^{-1}  \left( \frac{\hbar}{i} \nabla -  {\bf Q}  {\bf A} \right) + V  - E_k \right]  \Psi_k({\bf q})   \label{eq:Schroed3} 
\end{eqnarray}}

For $N=1$, to avoid the singularity of (\ref{eq:ChangeOfVariables}) at $V'({q}'_{sk})= 0$, we start the integration of (\ref{eq:ChangeOfVariables}) with the slope computed from L'Hôpital's rule 
\begin{equation}
    \frac{\partial {q}'_k}{\partial {q}}= \frac{\frac{\partial V}{\partial {q}}({q_{sk}})}{\frac{\omega^2}{m}  {q}_{sk}'} \ \ \ \ \ \ {\rm at} \ \ \ q'_{sk} = \sqrt{  \frac{2}{m \omega^2}E_k' } \label{eq:Hopital}
\end{equation}
implying $E_K = V({q_{sk}})$.
\label{th:ActionSolution}
\end{theorem}
Note that the intepretation and avoidance of singularities is current research.

For $N>1$, the system may first have to be decomposed into $N$ separate dynamics, using separation of variables or more generally the modular decomposition of nonlinear Hamiltonians recently derived in \cite{lohmiller2025contraction20naturalmetrics}.

Note that in contrast to solving the Schr\"odinger equation directly, the action and density calculations remain straightforward  in the nonlinear case. 
Theorem \ref{th:ActionSolution} avoids approximate perturbation theory \cite{Heisenberg1925, Schrodinger1926, Wilson1929}, using the WKB approximation \cite{Wentzel1926, Kramers1926,Brillouin1926} of partial differential equations, or semi-classical approximations \cite{landauquantum, VanVleck1928, Maslov1972, BerryMount1972Semiclassical}.



\section{Examples} \label{examples}

The following examples rewrite the Hamilton-Jacobi p.d.e. in the form (\ref{eq:HJscaled}), by first computing the change of variables ${\bf q}_k'({\bf q})$ from (\ref{eq:ChangeOfVariables}) in Theorem \ref{th:ActionSolution}.
A time-rescaling with $T_k$ then leads to a position independent propagated density (\ref{eq:densitydash}), even though the actions in the examples are not quadratic. This is a key difference to the semi-classical Van Vleck \cite{VanVleck1928} or Feynman \cite{Feynman} method which are only exact for a quadratic action, and thus cannot provide an exact result even for simple set-ups such as the double slit or the Aharonov-Bohm experiment.

The eigenwave of the time-independent Schr\"odinger equation (\ref{eq:Schroed3}) of the Hamiltonian (\ref{eq:HJscaled}) becomes a harmonic oscillator wave with a coordinate transformation from ${\bf q}$ to ${\bf q}_k'$.


We first illustrate the approach for the known wave of 3-dimensional hydrogen / Kepler oscillator, in essence re-deriving Example 3.10 of 
\cite{LohmillerSlotine2026QuantumWaves} using this unified framework. 
The approach is then applied to the quartic oscillator, a first case where only approximate solutions are known so far.
Both examples illustrate that the technique is straightforward and exact for nonlinear potential energies. 
They also illustrate that no Bohm quantum potential is needed in ${\bf q}_k'$ coordinates, where the classical propagated density computation yields a position independent $(T \Delta_{\bf M}) \phi(t')$. 


\Example{3D Coulomb or gravity potential}{\ \  \ \ Consider the Hamilton-Jacobi p.d.e. (\ref{eq:HJ}) of Theorem \ref{th:ActionSolution}
\begin{eqnarray}
- \frac{\partial \phi}{\partial t} \ = \ H \ = \ \frac{1}{2 m}  \frac{\partial \phi}{\partial {\bf q}}^T \frac{\partial \phi}{\partial {\bf q}} \ + \ \frac{G}{\sqrt{ {\bf q}^T {\bf q} }} \ - \ E_k    \nonumber 
\end{eqnarray}
with Cartesian position ${\bf q} = (q^1, q^2, q^3) \in \mathbb{R}^3$ \cite{LohmillerSlotine2026QuantumWaves}, constant mass $m$, energy offset $E_k$, and time $t \ge 0 $. $G$ corresponds to the Newtonian constant of gravitation scaled with the constant mass of the particle and the mass of the singularity, or alternatively to the Coulomb constant scaled with the constant charge of the particle and the charge of the singularity. 

Using (\ref{eq:ChangeOfVariables}) of Theorem \ref{th:ActionSolution}, the change of variables is
\begin{eqnarray}
    \frac{\partial {\bf q}'}{\partial {\bf q}}^T \frac{\partial {\bf q}'}{\partial {\bf q}} \ = \ T \ = \ \frac{\frac{G}{\sqrt{ {\bf q}^T {\bf q} }}  - E_k }{\frac{4 m \omega_k^2}{2}  {\bf q}^{'T} {\bf q}'  - E_k'} \ &=&  \ \frac{1}{4{\bf q}^{'T} {\bf q}' } \label{eq:quaternion}
\end{eqnarray}
where $ G = E_k' = \hbar \omega_k (k+\frac{4}{2}), \  k \in \mathbb{N}$ defines $E_k= \frac{4 m\ \omega_k^2}{2}$. Equation (\ref{eq:quaternion}) is the Eikonal equation for the 4-dimensional quaternions ${\bf q}' = (q^{'1}, ... q^{'4})$ \cite{Goldstein, LohmillerSlotine2026QuantumWaves}. The quaternion mapping implies $\mathbb{B} =  \{ \{ \uparrow, \downarrow \} \ \times k \ \in \mathbb{N}^\star  \}$  determined Kepler orbits \cite{Kepler} paths in ${\bf q}$ coordinates derived from a single path in ${\bf q}'$ \cite{LohmillerSlotine2026QuantumWaves}.


The 4 decoupled classical propagated actions $\phi(q^{'n}, t', q^{'n}_o)$ and densities $\rho(t')$ from $q^{'n}_o(q^{n}_o)$ are (\ref{eq:AnalyticAction}) which implies the kernel (\ref{eq:kernel}). The eigenwave (\ref{eq:EigenWave}) of Theorem \ref{th:ActionSolution} 
\begin{eqnarray}
    \Psi_{k}(q^n) \ &\propto&  H_k \left( { \sqrt{\frac{4 m\omega_k}{\hbar}}} q^{'n} (q^n) \right) e^{-\frac{4 m  \omega_k}{2  \hbar}(q^{n'}(q^n))^2} \nonumber
\end{eqnarray}
solves the original Schr\"odinger equation (\ref{eq:Schroed3}).
Note that this quaternion wave can be transferred in spherical coordinates using the 3-dimensional Hermite to Laguerre relation of Appendix A in \cite{Duru}. They were plotted in quaternion coordinates in \cite{LohmillerSlotine2026QuantumWaves}, confirming the correctness of the computation.

The approach in the current paper thus provides a more direct and standardized way to study this example, which requires time-rescaling.
As in \cite{LohmillerSlotine2026QuantumWaves}, and in contrast to the Feynman path integral \cite{Feynman} in the Duru-Kleinert propagator \cite{Duru}, no process noise is added to the classical path.}{Coulomb} 

We now turn to the simplest potential energy for which no exact solution of the Schr\"odinger equation is actually known. 


\Example{Quartic oscillator}{ \ \ Consider the Hamilton-Jacobi p.d.e. (\ref{eq:HJ}) of Theorem \ref{th:ActionSolution}
\begin{eqnarray}
- \frac{\partial \phi}{\partial t} \ = \ H \ = \  \frac{1}{2 m} \frac{\partial \phi}{\partial q}^2 \ + \ \frac{m}{2} \omega^2 q^4 \ - \ E_k  \nonumber 
\end{eqnarray}
with constant angular inertia $m$, constant angular frequency $\omega$, constant energy offset $E_k = \omega \hbar l$ with $l \in \mathbb{R}_{\ge 0}$, dimensionless position $q\in \mathbb{R}$, and time $t \ge 0 $. 

Using (\ref{eq:ChangeOfVariables}) of Theorem \ref{th:Hamilton} or \ref{th:ActionSolution} the change of variables is
\begin{eqnarray}
    T_k &=&  \frac{\frac{m}{2} \omega^2 q^4 - E_k}{\frac{m}{2} \omega^2 q_k^{'2} - E_k'} \ = \  \frac{ q^4  - 2l}{ q_k^{'2} - (2k+1)}  
    \nonumber \\
    q_k' &=& \int^q_0 \sqrt{ T_k } \  dq   \nonumber 
\end{eqnarray}
with $E_k' = \hbar \omega (k+\frac{1}{2}), \  k \in \mathbb{N}$ and where we use from now on the normalization $\hbar= m \omega$. 

We avoid a pole at $q'_{sk} = \pm \sqrt{  2k+1 }$ using  L'Hôpital's rule (\ref{eq:Hopital}),
\begin{equation}
    \frac{\partial {q}'_k}{\partial {q}} = \frac{2 q^3 } {  q'_{sk} }  \nonumber
\end{equation}
We choose $q_{sk}$ at $q'_{sk}  = -\sqrt{  2k+1 }$ such that the coordinate transformation passes through the origin $q' = q = 0$. This also assures that there is also no singularity at $q'_{sk} = \sqrt{  2k+1 }$, and it implies the dimensionless eigenenergy $ 2 l  = q_{sk}^4$.

As a result we get for $k=0, 1, 2, ...$ the change of variables in Figure \ref{fig:CoordQuartic} with the eigenergies $E_k = \omega \hbar l$. 
\begin{figure}
    \centering
    \includegraphics[width=4.2cm,height=7cm]{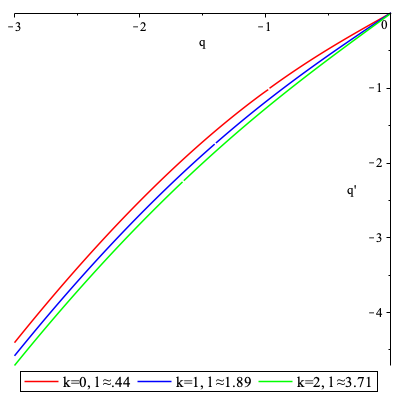}
    \caption{Coordinate transformation of the quartic oscillator} 
    \label{fig:CoordQuartic}
\end{figure}


The classical propagated action $\phi(q^{'}, t', q^{'}_o)$ and density $\rho(t')$ from $q^{'}_o(q_o)$ are (\ref{eq:AnalyticAction}) which implies the kernel (\ref{eq:kernel}). The eigenwaves (\ref{eq:EigenWave}) of Theorem \ref{th:ActionSolution} 
\begin{equation}
\Psi_k({q}) = \mathscr{H}_k({q}'_k({q})) = \frac{1}{\sqrt[4]{\pi}} \frac{1}{\sqrt{2^{k} k!}} \ H_{k} (q'(q)) e^{-\frac{1}{2} (q'(q))^{2}} \nonumber
\end{equation}
solves the original Schr\"odinger equation (\ref{eq:Schroed3}). The first three eigenwaves are illustrated in Figure \ref{fig:WaveQuartic}.
\begin{figure}
    \centering
    \includegraphics[scale=0.5]{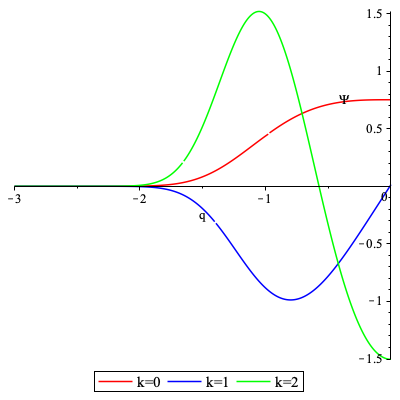}
    \caption{Eigenwaves of the quartic oscillator} 
    \label{fig:WaveQuartic}
\end{figure}
}{Quartic}

The last example illustrates that solving the Hamilton-Jacobi with time-rescaling is straightforward, whereas so far no direct solution of the  associated second-order Schr\"odinger equation was derived. It also illustrates why no Bohm-like quantum potential is needed in $q_k'$ coordinates, in which the classical action and density are computed.

\section{Summary} \label{concluding}

 This paper uses Theorem \ref{th:quantum}  to compute quantum waves exactly from the related classical action \cite{LohmillerSlotine2026QuantumWaves}. It focuses on the general nonlinear case, where the time-rescaling of Theorem \ref{th:Hamilton} assures that the propagated density from an initial ${\bf q}_o \veebar {\bf p}_o$ is independent of space, thus avoiding Bohm-like quantum potential terms. This is a key difference to the semi-classical approach of Van Vleck \cite{VanVleck1928} or the semi-classical version of Feynman \cite{Feynman}, which are only exact for a quadratic action.

Theorem \ref{th:ActionSolution} then shows that the time-rescaling in fact corresponds to a change of position variables (\ref{eq:ChangeOfVariables}) in the computed wave. In these new coordinates ${\bf q}_k'({\bf q})$, the eigenwaves $\Psi_k({\bf q})$ of the time-independent Schr\"odinger equation (\ref{eq:Schroed3}) of the Hamiltonian (\ref{eq:HJscaled}) are transformed into eigenwaves $\mathscr{H}_k({\bf q_k}'({\bf q}))$ of an harmonic oscillator. In a $1$-dimensional system, the resulting computation is immediate, since the nonlinear oscillator transforms into a harmonic oscillator (\ref{eq:EigenWave}), for which an exact analytic solution is already known. For higher-dimensional systems the same approach might be used, or the $N$-dimensional Hamiltonian system can be first decomposed into $N$ decoupled Hamiltonian dynamics.


In contrast to solving the Schr\"odinger equation (\ref{eq:Schroed}) directly, the exact action and density calculations thus extend naturally to the nonlinear case. While in general the solution of the nonlinear o.d.e. that defines the transformed variable may be numerical, this avoids the need for approximate perturbation theory \cite{Heisenberg1925, Schrodinger1926, Wilson1929}, the WKB approximation \cite{Wentzel1926} of partial differential equations, or semi-classical approximations \cite{landauquantum}.

The known 3D Coulomb wave \Ex{Coulomb} is used to illustate the theory. The wave solution of 
the quartic potential in \Ex{Quartic} is new, and shows that the approach is straightforward for general nonlinear potentials. 

Just as in \cite{LohmillerSlotine2026QuantumWaves}, the approach extends naturally to the relativistic context, with applications 
the subject of current research.




\ \




\normalem
\bibliographystyle{abbrv}
\bibliography{References}{}

\begin{thebibliography}{10}

\bibitem{beltrami1902ricerche}
E.~Beltrami.
\newblock {\em Ricerche di analisi applicata alla geometria}.
\newblock Vol. 1, Opere Matematiche, Milano, 1902.

\bibitem{BerryMount1972Semiclassical}
M.~V. Berry and K.~E. Mount.
\newblock Semiclassical approximations in wave mechanics.
\newblock {\em Reports on Progress in Physics}, 35:315--397, 1972.

\bibitem{Bohm1952I}
D.~Bohm.
\newblock A suggested interpretation of the quantum theory in terms of ``hidden'' variables, {I}.
\newblock {\em Physical Review}, 85(2):166--179, jul 1952.

\bibitem{Bohm1952II}
D.~Bohm.
\newblock A suggested interpretation of the quantum theory in terms of ``hidden'' variables, {II}.
\newblock {\em Physical Review}, 85(2):180--192, jul 1952.

\bibitem{Brillouin1926}
L.~Brillouin.
\newblock {La mécanique ondulatoire de Schrödinger: une méthode générale de résolution par approximations successives}.
\newblock {\em Comptes Rendus de l’Académie des Sciences (Paris)}, 183:24--26, 1926.

\bibitem{Duru}
I.~Duru and H.~Kleinert.
\newblock {Quantum Mechanics of H-atoms from path integrals}.
\newblock {\em {Fortschritte Phys.}}, 1982.

\bibitem{Euler}
L.~Euler.
\newblock {Principes généraux du mouvement des fluides}.
\newblock {\em Académie Royale des Sciences et des Belles Lettres}, 1755.

\bibitem{Feynman48}
R.~Feynman.
\newblock {Space-Time Approach to Non-Relativistic Quantum Mechanics}.
\newblock {\em Rev. Modern Physics}, 1948.

\bibitem{Feynman}
R.~Feynman and A.~Hibbs.
\newblock {\em Quantum Mechanics and Path Integrals}.
\newblock McGraw-Hill, 1965.

\bibitem{feynmanquantum1998}
R.~P. Feynman.
\newblock {\em Quantum Electrodynamics}.
\newblock Addison-Wesley, Reading, MA, 1998.

\bibitem{Goldstein}
H.~Goldstein.
\newblock {\em {Classical Mechanics}}.
\newblock Addison-Wesley, 1980.

\bibitem{Heisenberg1925}
W.~Heisenberg.
\newblock {Über quantentheoretische Umdeutung kinematischer und mechanischer Beziehungen}.
\newblock {\em Zeitschrift für Physik}, 33:879--893, 1925.

\bibitem{HenneauxTeitelboim1992}
M.~Henneaux and C.~Teitelboim.
\newblock {\em Quantization of Gauge Systems}.
\newblock Princeton University Press, 1992.

\bibitem{Kepler}
J.~Kepler.
\newblock {\em {Astronomia Nova}}.
\newblock Astronomia Nova, 1609.

\bibitem{Kleinert2009s}
H.~Kleinert.
\newblock {\em Path Integrals in Quantum Mechanics}.
\newblock World Scientific, 2009.

\bibitem{Kramers1926}
H.~A. Kramers.
\newblock {Wellenmechanik und halbzahlige Quantisierung}.
\newblock {\em Zeitschrift für Physik}, 39, 1926.

\bibitem{Lagrange}
J.~Lagrange.
\newblock {\em {Mécanique analytique}}.
\newblock Chez la veuve Desaint à Paris, 1788.

\bibitem{landauquantum}
L.~D. Landau and E.~M. Lifshitz.
\newblock {\em Quantum Mechanics: Non-Relativistic Theory}, volume~3 of {\em Course of Theoretical Physics}.
\newblock Pergamon Press, England, 1991.

\bibitem{laplace1799mecanique}
P.-S. Laplace.
\newblock {\em Mécanique Céleste}.
\newblock J.B.M. Duprat et al., Paris, 1799.

\bibitem{lohmiller2025contraction20naturalmetrics}
W.~Lohmiller and J.-J. Slotine.
\newblock Contraction 2.0 -- natural metrics in contraction analysis, 2025.

\bibitem{LohmillerSlotine2026BohmPotential}
W.~Lohmiller and J.-J. Slotine.
\newblock {Interpreting Bohm-like Quantum Potentials in "Computing Quantum Waves Exactly from Classical Action"}.
\newblock {\em arXiv preprint}, may 2026.

\bibitem{LohmillerSlotine2026QuantumWaves}
W.~Lohmiller and J.-J.~E. Slotine.
\newblock On computing quantum waves exactly from classical action.
\newblock {\em Proceedings of the Royal Society A}, 482(2336):20250413, 2026.

\bibitem{Lovelock}
D.~Lovelock and H.~Rund.
\newblock {\em Tensors, Differential Forms, and Variational Principles}.
\newblock Dover, 1989.

\bibitem{Madelung1927}
E.~Madelung.
\newblock {Quantentheorie in hydrodynamischer Form}.
\newblock {\em Zeitschrift f{\"u}r Physik}, 40(3-4), 1927.

\bibitem{Maslov1972}
V.~P. Maslov.
\newblock {\em Théorie des perturbations et méthodes asymptotiques}.
\newblock Éditions Mir, 1972.

\bibitem{Nash}
J.~Nash.
\newblock {$C^1$ Isometric Embeddings}.
\newblock {\em {Annals of Mathematics. 2nd Series, 60 (3)}}, 1954.

\bibitem{Nielsen1988PathIntegralQuantizeSpin}
H.~B. Nielsen and D.~Rohrlich.
\newblock A path integral to quantize spin.
\newblock {\em Nuclear Physics B}, 299:471--483, 1988.

\bibitem{Schrodinger1926}
E.~Schrödinger.
\newblock {Quantisierung als Eigenwertproblem}.
\newblock {\em Annalen der Physik}, 79:361--376, 1926.

\bibitem{Schulman1968PathIntegralSpin}
L.~S. Schulman.
\newblock A path integral for spin.
\newblock {\em Physical Review}, 176:1558--1569, 1968.

\bibitem{VanVleck1928}
J.~H.~V. Vleck.
\newblock The correspondence principle in the statistical interpretation of quantum mechanics.
\newblock {\em P.N.A.S}, 14(2):178--188, 1928.

\bibitem{Wentzel1926}
G.~Wentzel.
\newblock {Eine Verallgemeinerung der Quantenbedingungen für die Zwecke der Wellenmechanik}.
\newblock {\em Zeitschrift für Physik}, 38:518--529, 1926.

\bibitem{Wilson1929}
A.~Wilson.
\newblock Perturbation theory in quantum mechanics.
\newblock {\em Proc. Roy. Soc. A}, 1929.

\end{thebibliography}


\end{document}